\documentclass[aps,prl,preprint,superscriptaddress]{revtex4-1}
\usepackage{graphicx}
\usepackage{dcolumn}
\usepackage{bm}
\usepackage[usenames]{color}
\usepackage{amsmath}
\usepackage{amssymb}
\usepackage{latexsym}

\usepackage[left]{lineno}

\begin{document}

\title{Co-appearance of superconductivity and ferromagnetism in a Ca$_2$RuO$_4$ nanofilm crystal
}

\author{Hiroyoshi Nobukane}
\email{nobukane@sci.hokudai.ac.jp}
\thanks{Corresponding author}
\affiliation{Department of Physics, Hokkaido University, Sapporo, 060-0810, Japan}
\affiliation{Center of Education and Research for Topological Science and Technology, Hokkaido University, Sapporo, 060-8628, Japan}

\author{Kosei Yanagihara}
\affiliation{Department of Physics, Hokkaido University, Sapporo, 060-0810, Japan}

\author{Yuji Kunisada}
\affiliation{Center for Advanced Research of Energy and Materials, Hokkaido University, Sapporo, 060-0828, Japan}

\author{Yunito Ogasawara}
\affiliation{Department of Physics, Hokkaido University, Sapporo, 060-0810, Japan}

\author{Kakeru Isono}
\affiliation{Department of Physics, Hokkaido University, Sapporo, 060-0810, Japan}

\author{Kazushige Nomura}
\affiliation{Department of Physics, Hokkaido University, Sapporo, 060-0810, Japan}

\author{Keita Tanahashi}
\affiliation{Center for Advanced Research of Energy and Materials, Hokkaido University, Sapporo, 060-0828, Japan}

\author{Takahiro Nomura}
\affiliation{Center for Advanced Research of Energy and Materials, Hokkaido University, Sapporo, 060-0828, Japan}

\author{Tomohiro Akiyama}
\affiliation{Center of Education and Research for Topological Science and Technology, Hokkaido University, Sapporo, 060-8628, Japan}
\affiliation{Center for Advanced Research of Energy and Materials, Hokkaido University, Sapporo, 060-0828, Japan}

\author{Satoshi Tanda}
\affiliation{Center of Education and Research for Topological Science and Technology, Hokkaido University, Sapporo, 060-8628, Japan}
\affiliation{Department of Applied Physics, Hokkaido University, Sapporo 060-8628, Japan}

\date{\today}

\begin{abstract}
\textbf{
By tuning the physical and chemical pressures of layered perovskite materials we can realize the quantum states of both superconductors and insulators.
By reducing the thickness of a layered crystal to a nanometer level, a nanofilm crystal can provide novel quantum states that have not previously been found in bulk crystals.
Here we report the realization of high-temperature superconductivity in Ca$_2$RuO$_4$ nanofilm single crystals.
Ca$_2$RuO$_4$ thin film with the highest transition temperature $T_c$ (midpoint) of 64~K exhibits zero resistance in electric transport measurements.
The superconducting critical current exhibited a logarithmic dependence on temperature and was enhanced by an external magnetic field.
Magnetic measurements revealed a ferromagnetic transition at 180~K and diamagnetic magnetization due to superconductivity.
Our results suggest the co-appearance of superconductivity and ferromagnetism in Ca$_2$RuO$_4$ nanofilm crystals.
We also found that the induced bias current and the tuned film thickness caused a superconductor-insulator transition.
The fabrication of micro-nanocrystals made of layered material enables us to discuss rich superconducting phenomena in ruthenates.
}
\end{abstract}

\maketitle

\noindent
\subsection{Introduction}
The search for high-temperature (high-$T_c$) superconductors is a fascinating topic in condensed matter physics.
It is widely believed that high-$T_c$ superconductivity in cuprates emerges from doped Mott insulators~\cite{Bednorz}.
Recently, $4d$ and $5d$ transition metal oxides with a layer perovskite structure have attracted much attention because the possibility of the emergence of high-$T_c$ superconductivity has been recognized in several studies~\cite{Chaloupka,Feng,Kim}.
Indeed, monolayer films in iron pnictides indicate the enhancement of $T_c$ to above 100~K~\cite{Jia}.
By tuning the film thickness in the monolayer to the nanometer range, transition metal dichalcogenides realize an exotic ground state different from that of bulk crystals due to a negative pressure effect~\cite{Tsen,Goli}.
Thus, the layered nanoscale film crystals play a key role when we explore the emergence of high-$T_c$ superconductivity in layered perovskite $4d$ and $5d$ transition metal oxides, which may allow us to detect superconductivity on mesoscopic scales.

For unconventional superconductors in cuprates~\cite{Lake,Chang_PRB}, ruthenates~\cite{Maeno}, iron pnictides~\cite{Kamihara}, organic~\cite{Oka} and heavy-fermion materials~\cite{Saxena,Aoki}, it is important to reveal the interplay between superconductivity and magnetism.
Magnetic interactions are closely related to the mechanism of superconductivity, which attracts electrons towards each other.
The antiferromagnetic correlations in high-$T_c$ cuprate superconductors lead to spin-singlet $d$-wave pairing states.
In contrast, ferromagnetic correlations favour spin-triplet pairing states.
The superfluidity of $^3$He~\cite{Volovik} is a leading physical system in which spin-triplet pairing is realized at very low temperatures.
The coexistence of superconductivity and ferromagnetism in uranium compounds~\cite{Saxena,Aoki} has attracted much attention.
Layered perovskite Sr$_2$RuO$_4$ ($T_c=1.5$~K) is a leading candidate for a spin-triplet and chiral $p$-wave superconductor in quasi-two-dimensional electron systems~\cite{Maeno_RMP,Nobukane_PRB}.
In rutheno-cuprate superconductors~\cite{Felner,Tallon}, superconductivity and ferromagnetism appear to coexist in different layers of layered perovskite structures where the superconductivity is confined to the antiferromagnetic CuO$_2$ planes and is not caused by the spin-triplet pairing associated with the ferromagnetism of the RuO$_2$ planes.
However, high-$T_c$ ferromagnetic superconductors have yet to be found, while the spin-triplet superconductivity and superfluidity that have been reported were realized at very low temperatures.
Here we report the observation of high-$T_c$ superconductivity related to ferromagnetism in Ca$_2$RuO$_4$ crystals with a nanoscale thickness.
A Ca$_2$RuO$_4$ thin film exhibits zero resistance and diamagnetic magnetization at high temperature as evidence of superconductivity.
Intriguingly, the enhancement of the critical current and the diamagnetic component for the applied magnetic field reveal the co-appearance of superconductivity and ferromagnetism.
Moreover, we also found that the induced bias current and the tuned film thickness cause a superconductor-insulator (SI) transition.
The superconductivity in Ca$_2$RuO$_4$ thin film crystals is expected to be robust against external magnetic fields, and play an important role when applied to a topological quantum computation~\cite{Nayak_RMP}.

The electronic states in layered perovskite $A_2$RuO$_4$ have various phases that depend on metallic elements $A$~\cite{nakatsuji_jpsj}.
The ruthenate Sr$_2$RuO$_4$ is a superconductor below a transition temperature $T_c$ of 1.5~K~\cite{Maeno}.
The substitution of Sr with Ca chemically pressurises the perovskite and causes the lattice distortion of RuO$_6$ octahedra, which strongly modifies the electronic structures.
As a consequence, a ruthenate where $A=$Ca$_{2-x}$Sr$_x$ becomes a paramagnetic metal or an antiferromagnetic metal depending on the doping rate $x$~\cite{nakatsuji_prl,Friedt,Gorelov}.
At $x=0$, the large degree of lattice distortion turns Ca$_2$RuO$_4$ into a band-Mott insulator with an antiferromagnetic correlation~\cite{sutter,das}.
In contrast, an electron doping system where $A=$Ca$_{2-x}$La$_x$~\cite{Cao} and thin films~\cite{Miao} shows a transition to ferromagnetic ordering.
Experimental control of the distortions in Ca$_2$RuO$_4$ drives the transition from an insulator to a metal by applying hydrostatic and uniaxial pressure~\cite{nakamura_prb,Braden,nakamura_jpsj,taniguchi} and by applying electric fields~\cite{nakamura_srep} and bias current~\cite{sow}.
Such physical pressures are considered to release the distortion of octahedra.
Interestingly, an ac susceptibility measurement performed in bulk Ca$_2$RuO$_4$ in the 9 $\sim$ 14 GPa range~\cite{Alireza} showed a transition to the superconducting phase below 0.4 K.
Pressurization plays a key role in modifying the electric states in the ruthenate family.
A third way of controlling the lattice distortion might be to tune the dimensionality of Ca$_2$RuO$_4$ crystal.
Recent studies have reported the effects of negative pressure on lattice distortion when the thickness of the exfoliated films decreases to the nanometer range~\cite{Tsen,Goli}.
We have already reported the novel transport properties of Sr$_2$RuO$_4$ nanofilm single crystals~\cite{nobukane_srep}.
In this article, we describe superconducting phenomena at high temperature induced by ferromagnetism in thin films of Ca$_2$RuO$_4$.

\subsection{Results and Discussion}
Figure~\ref{figure1}a shows the powder X-ray diffraction (XRD) patterns for Ca$_2$RuO$_4$.
All the peaks can be well indexed with an orthorhombically distorted K$_2$NiF$_4$ structure characterized by a long $c$-axis (space group: \textit{L-Pbca}).
Minor peaks corresponding to the CaRuO$_3$ ($\ast$) and Ca$_2$RuO$_4$ ($\sharp$) (\textit{S-Pbca}) phases were also observed.
The estimated CaRuO$_3$ phase was $\sim 5\%$ whereas the amount of Ca$_2$RuO$_4$ (\textit{S-Pbca}) was $\sim 6\%$.
The lattice parameters $a=5.350(2)$~\AA\,, $b=5.343(3)$~\AA\,, and $c=12.278(5)$~\AA\, were obtained, which are close to those of the L phase reported for Ca$_2$RuO$_4$ under pressure~\cite{Braden} and Ca$_{2-x}$Sr$_x$RuO$_4$~\cite{Friedt}. We found that the Ca$_2$RuO$_4$ nanofilm crystals were L phase at room temperature and ambient pressure.
The impurity phase CaRuO$_3$ is a paramagnetic metal down to low temperature in bulk~\cite{Cao_113}.
Although the behaviour of a ferromagnetic metal in CaRuO$_3$ films has been reported~\cite{Rana}, they are different from the electric and magnetic properties of our samples.
CaRuO$_3$ films exhibit no diamagnetism~\cite{Rana}.
In addition, we were able to distinguish layered Ca$_2$RuO$_4$ and CaRuO$_3$ (cubic) by observing the crystal shape using a scanning electron microscope.
We measured selected Ca$_2$RuO$_4$ nanofilm crystals as shown in the inset of Fig.~\ref{figure1}c.

Figure~\ref{figure1}b shows the temperature dependence of the longitudinal resistivity $\rho$ of Ca$_2$RuO$_4$ nanofilm crystals and the dependence of magnetic susceptibility $M/H$ on temperature at 10~Oe for powders weighing 5.2~mg (sample A).
The resistivity $\rho=10^{-4} - 10^{-2}~\Omega~\mathrm{cm}$ at 290~K of the samples is much lower than that in the $ab$-plane $\rho_{ab}\sim 6~\Omega \cdot$cm in bulk Ca$_2$RuO$_4$ (S phase) and is more like the resistivity of the L phase under hydrostatic pressure above 0.5 GPa~\cite{nakamura_prb}.
This behaviour is consistent with the XRD result obtained with thin film crystals.
We found a drop in the resistivity to zero and a diamagnetic component for samples with a broad transition behaviour of $\rho$ and $M/H$.
Magnetization data are discussed in detail in Fig.~\ref{figure2}.
In Fig.~\ref{figure1}c, the dependence of resistivity on temperature was not suppressed when a magnetic field was applied parallel to the $c$ axis.
Moreover, Fig.~\ref{figure1}d shows the current-voltage ($I-V$) characteristics of sample 3 for several magnetic fields at 0.53~K, which are vertically shifted for clarity.
In $B=0$, the result clearly shows supercurrents of $|I_{c}|< 13$~nA for the low bias current region, which strongly suggests the presence of superconductivity in a Ca$_2$RuO$_4$ thin film.
The supercurrent was observed at 5~K even in 17~T (see Supplementary Fig.~S2e).
The superconductivity in the measured samples is robust for the applied magnetic field.
Interestingly, the supercurrent increases as the magnetic field increases.
Our results show that superconductivity is enhanced by an applied magnetic field, although the external magnetic field generally suppresses superconductivity in conventional superconductors.
We also observed hysteresis with the supercurrent in the $I-V$ curves for sample 3 as shown in the inset of Fig.~\ref{figure1}e.
The hysteresis behaviour reveals a superconducting phase property because the hysteresis is a characteristic feature of small Josephson junctions, Josephson junction arrays and superconducting nanowires~\cite{zant_phaseslip}.
The result suggests that a number of superconducting domains emerge locally at low temperature and that such domains interact with one another through the Josephson coupling.
Similar behaviour also occurs in a phase-slip in superconducting nanowires~\cite{Tinkham_book}.
In Fig.~\ref{figure1}e, the critical current $I_c$ is enhanced at low temperatures.
Although the Josephson critical current in conventional superconductors is described by the Ambegaokar-Baratoff relation~\cite{Ambegaokar}, the $I_c$ anomaly cannot be explained by this relation (the dotted line in Fig.~\ref{figure1}e).
Considering that the applied magnetic field enhances rather than suppresses the superconductivity, we use that Josephson critical current relation between chiral $p$-wave superconductors~\cite{Barash,Asano}, $I_{c}(T)=aI_{c0}~\mathrm{ln}\left( \frac{b \Delta}{T} \right)\mathrm{sin}\Phi$, where $a$ and $b$ are constants of the order of unity, and the phase difference $\Phi$.
The fitting result is shown in Fig.~\ref{figure1}e, where we fixed $\mathrm{sin}\Phi =1$.
The blue solid line with the logarithmic dependence fits nicely with our experimental data.
This result suggests that the superconductivity of Ca$_2$RuO$_4$ nanofilm crystals realizes the chiral $p$-wave state.
The deviation of the data from the fitting curve below 10~K seems to be saturation of $I_c$, which changes with the transparency of the junction~\cite{Barash}.
Moreover, in Ref.~\cite{Asano}, a suppression of $I_c$ has been reported when the angular momentum of Cooper pairs in two chiral domains align in anti-parallel.
We emphasize that the critical current is larger than that in $s$-wave junctions.
We note that the broad transition behaviour of $\rho$ and $M/H$ is accentuated in two-dimensional (2D) superconductivity.
A similar broad transition with $\Delta T=30-40$~K has been reported in YBa$_2$Cu$_3$O$_{7-\delta}$ films~\cite{matsuda,Leng} and FeSe film~\cite{Shiogai}.
The Berezinskii, Kosterlitz and Thouless~\cite{Berezinskii,Kosterlitz_Thouless} (BKT) transition for a 2D superconductor was observed in our samples (See Supplementary Text and Fig.~S2).
Our samples show two onset temperatures $T_{c~\mathrm{High}}^{\mathrm{on}}$ of $80-100$~K (light green) and $T_{c~\mathrm{Low}}^{\mathrm{on}}$ of $\sim 50$~K (green).
The highest onset temperature 96~K was observed.
The $T_{c~\mathrm{High}}^{\mathrm{on}}$ value is close to the temperature at which diamagnetism begins to appear in Fig.~\ref{figure1}b, and the $T_{c~\mathrm{Low}}^{\mathrm{on}}$ value corresponds to the temperature at which the diamagnetic component is at its maximum.

To clarify the diamagnetism of the superconductivity, we performed magnetic measurements for powders weighing 5.2 mg (sample A) and 2.6~mg (sample B).
Figures~\ref{figure2}a and ~\ref{figure2}b show the dependence of magnetic susceptibility on temperature for the standard field-cooling (FC) process.
The susceptibility of $\sim 5.0 \times 10^{-5}$ emu/cm$^3$ above $T_{\mathrm{Curie}}=$180~K is consistent with that reported for bulk Ca$_2$RuO$_4$~\cite{nakatsuji_jpsj,taniguchi,sow}.
Below $T_{\mathrm{Curie}}$, spontaneous magnetization increases sharply as shown in the inset of Fig.~\ref{figure2}a.
We also observed a magnetic hysteresis loop below $T_{\mathrm{Curie}}$ (see Supplementary Fig.~S3), which indicates a ferromagnetic state.
The non-saturating behaviour of the magnetization in high fields reflects the weak itinerant ferromagnetism of Ca$_2$RuO$_4$ thin films.
In reports containing similar results to ours, the itinerant ferromagnetism in Ca$_2$RuO$_4$ appears in the ``L" phase of bulk Ca$_2$RuO$_4$~\cite{nakatsuji_jpsj}, La-doped Ca$_2$RuO$_4$~\cite{Cao} and epitaxial thin films~\cite{Miao}.
Surprisingly, a diamagnetic component in our nanofilm crystals appears as shown in Figs.~\ref{figure2}a and~\ref{figure2}b.
We observed the diamagnetic behaviour even in 7~T as shown in Fig.~\ref{figure1}c.
This means that superconductivity is not eliminated by a magnetic field of 7~T, which is consistent with transport results.
Recently, current-induced diamagnetism has been reported in bulk Ca$_2$RuO$_4$~\cite{sow}.
However, the result we obtained in a low field is sufficiently larger than the value of the diamagnetism in Ref.~\cite{sow}.
Here we performed two types of subtractions to estimate the diamagnetic component.
We first isolated the diamagnetic component $(\Delta M/H)_{\mathrm{min}}$ by subtracting the peak value of the magnetic susceptibility at $\sim$130~K from data points for each field as shown in Figs.~\ref{figure2}c and ~\ref{figure2}d.
The diamagnetism value was $\left(4\pi \Delta M/H\right)_{\mathrm{min}} \sim -0.09~\mathrm{emu/cm^3}$ at 10 Oe.
We also performed an analysis using a fitted curve with temperature dependent magnetization in the itinerant ferromagnetism (see Supplementary Fig.~S4).
In ferromagnetic superconductors, the itinerant ferromagnetic fit has been subtracted off to reveal the superconducting part~\cite{Aoki,Paulsen}.
The diamagnetism for our samples lies within the range of $(4\pi \Delta M/H)_{\mathrm{min}}= -0.09~\mathrm{emu/cm^3} < 4\pi \Delta M/H <(4\pi \Delta M/H)_{\mathrm{max}}=-0.32~\mathrm{emu/cm^3}$ at 10~Oe.
Here we note that the diamagnetic behaviour is qualitatively the same regardless of the subtraction approach.
The diamagnetism is reproduced in measured samples.
Interestingly, the diamagnetization at 10~Oe is about 500-1000 times stronger than that in other non-superconducting materials such as highly oriented pyrolytic graphite and bismuth.
There are no phenomena that show such giant diamagnetism other than superconductivity.
Therefore, this result is evidence that Ca$_2$RuO$_4$ nanofilm crystals exhibit high-$T_c$ superconductivity.
We note that a giant diamagnetic response above $T_c$ has been reported in cuprates~\cite{Li} and FeSe~\cite{Kasahara}.
Their experiments reveal giant diamagnetism as a precursor to superconductivity and pseudogap formation.
The diamagnetism response above $T_c$ for our samples may be considered the signature of preformed Cooper pairing (see Supplementary Information).

Next, to realize a ferromagnetic order, we cooled a powder of sample B at 5000 Oe from 190~K to 145~K ($<T_{\mathrm{Curie}}=180$~K), and then measured the temperature dependence of the magnetization in fields of 5, 8, 10, 12, 15 and 100 Oe from 145 to 2~K (see Supplementary Fig.~S5).
We raised the temperature to 190~K ($>T_{\mathrm{Curie}}=180$~K) before performing each measurement.
The magnetic susceptibility result for sample B is shown in Fig.~\ref{figure2}e.
In Fig.~\ref{figure2}f, the diamagnetic component $\left(4\pi \Delta M/H\right)_{\mathrm{min}}$ is plotted by subtracting the peak value of the susceptibility, which is similar to the behaviour of sample B in Fig.~\ref{figure2}d.
Moreover, to compare the diamagnetism values, Fig.~\ref{figure2}g shows the result of an FC measurement from 190 to 2~K at 5~Oe and the result for an FC measurement from 145 to 2~K at 5~Oe after cooling from 190 to 145~K while applying a magnetic field of 5000~Oe.
Interestingly, we found that the diamagnetic components were enhanced by ferromagnetic ordering in sample B.
This means that superconductivity and ferromagnetism coexisted.
The results of the electric transport and magnetic measurements suggest that the superconductivity emerges locally below $T_{\mathrm{Curie}}$ and that a drop in the resistance to zero is detected through the transport when the superconducting domain increases in size in the presence of a ferromagnetic correlation in Ca$_2$RuO$_4$ thin films.
We note that our results for high-$T_c$ superconductivity in nanofilm crystals are inconsistent with the transition to the superconducting phase at $T_c=$ 0.4 K in bulk Ca$_2$RuO$_4$ under high pressures exceeding 9 GPa~\cite{Alireza}.
We believe this to be due to the difference in the temperature at which the ferromagnetic phase appears.
In nanofilm crystals, ferromagnetic ordering appears below 180~K, whereas bulk Ca$_2$RuO$_4$ exhibits ferromagnetism below 30~K~\cite{Alireza}.

The effects of quantum fluctuations in two dimensions would be more dominant in a Ca$_2$RuO$_4$ film.
In Fig.~\ref{figure3}a, we plot the sheet resistance $R_{\Box/layer}=\rho / d$ of sample 2 as a function of temperature for different bias currents $I$, where the interlayer distance of Ca$_2$RuO$_4$ $d=6.13$ \AA\,.
Correspondingly, $R_{\Box/layer}$ is the resistance per square per RuO$_2$ layer.
In the low temperature range from 4.2 to 20~K, for $I<I_{\mathrm{QCP1}}=100$ nA, $R_{\Box/layer}$ decreases with decreases in temperature, while $R_{\Box/layer}$ increases with decreases in temperature for $I>I_{\mathrm{QCP1}}$.
Recently, by tuning the DC current, a current-induced metal-insulator transition has been realized in bulk Ca$_2$RuO$_4$~\cite{sow}.
The electronic states in Ca$_2$RuO$_4$ are sensitive to DC current.
Our results suggest that the application of a DC current induces a transition from 2D superconductivity to an insulator.
The critical current density value for high-$T_c$ thin film superconductors is $j_c > 10^4$ A/cm$^2$~~\cite{Talantsev}.
At $I_{\mathrm{QCP1}}=100$ nA in Fig.~\ref{figure3}a, the current density estimated as $j \approx 8.3 \times 10^2$ A/cm$^2$ is much smaller than $j_c$  (see Supplementary Table S1 and Text).
We note that a current-driven SI transition has been reported in cuprate thin film superconductors~\cite{Santos}.
Figure~\ref{figure3}b shows the bias current dependence of $R_{\Box/layer}$ for different temperatures from 4.2 to 20~K.
The crossing point is $I_{\mathrm{QCP1}}=100~\mathrm{nA}$, $R_{\mathrm{QCP1}}=16.5~\mathrm{k}\Omega \sim \frac{8}{\pi}\frac{h}{4e^2}$.
Near the SI transition point, the sheet resistance should obey
$R_{\Box}(I, T)=\frac{h}{4e^2} f \left[\frac{c_{0}(I-I_c)}{T^{1/z\nu}}\right]$, where $c_0$ is a constant making the argument dimensionless, and $I_c$ is a current at a quantum critical point.
The dynamical and correlation-length exponents are denoted by $z$ and $\nu$, respectively~\cite{fisher}.
To determine the critical exponent $z\nu$ from the experimental results, we use $(dR_{\Box/layer}/dI)_{I=I_c}$ at a fixed $T$, i.e., $(dR_{\Box/layer}/dI)_{I=I_c}=\frac{c_{0}h}{4e^2}T^{-1/z\nu} \tilde{R'}(0)$, where $\tilde{R'}(0)$ is finite.
The critical exponent $z\nu$ at low temperature was found to be 1.5 by plotting  $(dR_{\Box/layer}/dI)_{I=I_\mathrm{QCP1}})$ as a function of $T^{-1}$ as shown in ~Fig.~\ref{figure3}d.
The inset in Fig.~\ref{figure3}b shows the scaling behaviour of the sheet resistance, where we use $I_{\mathrm{QCP1}}=100$~nA and $z\nu=1.5$.
The results show that all of the resistance data below 20 K fall on a universal curve.
Similarly, we performed a scaling analysis in a high temperature range of 30 to 45~K.
Figure~\ref{figure3}c and its inset show the bias current dependence of $R_{\Box/layer}$ for different temperatures and the universal curve at $z\nu=0.64$, where $I_{\mathrm{QCP2}}=1960$~nA and $R_{\mathrm{QCP2}}=38.1~\mathrm{k}\Omega \sim 6\frac{h}{4e^2}$ were used.
We observed two critical exponents as shown in Fig.~\ref{figure3}d.
Our result is very similar to two-stage quantum critical behaviour and the critical exponents reported in a magnetic-field-induced SI transition~\cite{Lesueur,Popovic}.
Indeed, the critical exponent for the scaling analysis estimated from $I-V$ characteristics with $I^{*}_{\mathrm{QCP}}=\pm 140$~nA for sample 4 (inset in Fig.~\ref{figure3}e) was $z\nu = 0.68$ at low temperature as shown in Fig.~\ref{figure3}e, which is consistent with that in the high-temperature region for sample 2.
Sample 2 contains more disorder or inhomogeneity than sample 4 from resistivity results.
Therefore, quantum phase fluctuations seems to play an important role at low temperature.

Now we discuss the two critical exponents in a 2D superconductor.
In general, the behaviour in a quantum (or classical) critical phase transition is classified in universality classes that depend on the properties of the system such as its symmetries and dimensionality.
The critical exponents for disordered systems must satisfy the Harris criterion, which is represented by $\nu \ge 2/d$, where $d$ is the spatial dimensionality~\cite{fisher}.
The dynamical exponent is assumed to be $z=1$ due to the long-range Coulomb interaction between charges.
The critical exponent of $\nu\sim 2/3$ corresponds to the (2+1)D \textit{XY} model for a 2D superconductor~\cite{Lesueur}.
In contrast, the critical exponent $\nu=3/2$ suggests that the SI transition is explained by a Bose glass in two dimensions.
Previous papers have reported quantitatively similar values for the critical exponent in InO$_{x}$~\cite{Paalanen}, Nd$_{2-x}$Ce$_x$CuO$_4$~\cite{Tanda}, $\alpha-$MoGe~\cite{Yazdani} and La$_{2-x}$Sr$_{x}$CuO$_4$~\cite{Bollinger} films.
The critical value of $R_{\mathrm{QCP1}}$ is very close to the universal resistance $\frac{8}{\pi}\frac{h}{4e^2}$ predicted by the superconductor to Mott-insulator transition~\cite{Fisher_Girvin}.
At low temperature, the contribution of disorders caused by quantum fluctuations is enhanced and may appear as intrinsic inhomogeneities on a micro-nanoscale with the coexistence of superconducting and non-superconducting domains in strongly correlated 2D superconductors.
We note that, at present, we cannot determine the universality class for our samples because classical percolation ($\nu=4/3$) and quantum percolation ($\nu=7/3$) models have been reported as the expected critical exponents of $\nu \ge 1$~\cite{Steiner}.
We observed hysteresis with the supercurrent in the $I-V$ curves for sample 2 as shown in the inset in Fig.~\ref{figure3}d.
This agrees well with the existence of the arrays of superconducting domains (Fig.~\ref{figure3}d) explained by $z\nu =3/2$ for disordered (strongly correlated) superconductors.
One reason why a DC current-induced SI transition occurs is that the coupling between the superconducting domains is broken as the DC current is increased.
Thus, a current-driven SI transition occurs.
Another reason is that an SI transition is caused by driving the ferromagnetic superconducting domains (walls) by applying a DC current.

Figure~\ref{figure3}f shows the dependence of the resistivity $\rho$ on temperature for different thicknesses of Ca$_2$RuO$_4$ single crystal.
In general, the resistivity in conventional metals becomes an intrinsic property irrespective of their size and shape.
As the sample thickness decreases, the resistivity in thin films tends to increase due to the contribution of dislocations and impurities.
Our result in Fig.~\ref{figure3}f is the complete opposite of the above case.
The resistivity of Ca$_2$RuO$_4$ thin flakes is thickness dependent and decreases as the thickness decreases to the nanometer range.
Surprisingly, the resistivity for $10-30$~nm thick samples decreases below $T_{c}^{\mathrm{on}}$, while sample 7 with a thickness of 300~nm exhibits insulating behaviour.
The critical thickness is 50~nm.
We discuss the reasons for the different transport properties of samples with approximately the same thickness of $\sim 10$~nm (see Supplementary Table S2 and Text).
By reducing the thickness to the nanometer range, we succeeded in controlling the electronic states from an insulator to a superconductor.

Below, we explain the reasons for the superconducting behaviour in a Ca$_2$RuO$_4$ thin film.
As shown in Fig.~\ref{figure4}, the electronic properties in a Ca$_2$RuO$_4$ thin film are qualitatively different from those in a Mott insulator Ca$_2$RuO$_4$.
Previous studies have partially explained the reasons for the difference.
Octahedra consisting of RuO$_6$ in bulk Ca$_2$RuO$_4$ have three types of distortion (flattening, tilting, and rotating)~\cite{Friedt,nakamura_prb,Braden,nakamura_jpsj,taniguchi}.
The application of physical or chemical pressure releases the distortion, which  leads to a transition from the Mott insulating phase to the ferromagnetic metallic phase~\cite{Friedt,nakamura_prb,Braden,nakamura_jpsj,taniguchi}.
We discuss why the thin films are metallic rather than insulating.
When the thickness of a layered thin film decreases to the nanometer range~\cite{Tsen,Goli,nobukane_srep}, the effects of negative pressure can release the distortion of the octahedra.
The results of our first-principle calculation in Fig.~\ref{figure4} partially support this argument.
The figure summarizes the relationship between the number of RuO$_2$ layers along the $c$ axis and the tilting angle of the octahedra from the $c$ axis (see Supplementary Table S3 and Text).
The tilting angle decreases and the distance along the $c$ axis increases as the number of layers decreases.
The calculation also suggests that there is a uniform ferromagnetic metal phase in a Ca$_2$RuO$_4$ monolayer in the absence of flattening distortion.
Actually, a change of a few percent in the lattice constant and the distortion of the octahedra affect the electric and magnetic properties in transition metal oxides with a layered perovskite structure such as RuO$_6$~\cite{Miao,Dietl,Steppke}, CuO$_6$~\cite{Thio} and IrO$_6$~\cite{Crawford,Korneta}.
Therefore, the suppression of the lattice distortion under a negative pressure is responsible for the superconducting properties in a thin film single crystal.

Finally, we discuss the pairing symmetry of superconductivity in Ca$_2$RuO$_4$ flakes.
We revealed that the temperature dependence of $I_c$ can be understood by the fitting in chiral $p$-wave superconductors.
The critical current and diamagnetism was enhanced by the applied magnetic field.
These results suggest the possibility of spin-triplet Cooper pairing in Ca$_2$RuO$_4$.
We must undertake further experimental studies to clarify the mechanism of the pairing symmetry and unconventional vortices~\cite{tafari,Bert,Stolyarov} using STM and a scanning SQUID microscope.

\subsection{Conclusion}

\vspace{0.2cm}
In conclusion, we have studied electric transport and magnetic properties in a Ca$_2$RuO$_4$ thin film.
Our transport experiments show the logarithmic dependence of the $I_c$ on temperature, and the enhancement of $I_c$ under an external magnetic field applied perpendicular to the conducting plane.
Magnetic measurements revealed the ferromagnetic transition and the diamagnetism caused by superconductivity.
We observed current-induced and film-tuned SI transitions in a Ca$_2$RuO$_4$ thin film, which are phenomena unique to 2D superconductors.

\subsection{Methods}

To obtain micro-nanoscale Ca$_2$RuO$_4$ single crystals, we synthesized Ca$_2$RuO$_4$ crystals with a solid phase reaction.
We prepared CaCO$_3$ (99.99$\%$, Kojundo Chem.) and RuO$_2$ (99.9$\%$, Kojundo Chem.) powders.
The mixed powder was heated at 1070 $^\circ$C for 60 hours.
The mixture was then cooled gradually to room temperature.
We repeated the heating and cooling process several times.
The Ca$_2$RuO$_4$ crystal structure was analysed by using XRD (Rigaku Miniflex600) with Cu $K\alpha$ radiation.
The samples were dispersed in dichloroethane by sonication and deposited on a SiO$_2$(300~nm)/Si substrate.
We then fabricated gold electrodes using standard electron beam lithography methods.
The sample thickness was determined from scanning electron micrographs obtained with a sample holder tilt of 70 degrees.

We performed electric transport measurements with the four-terminal method.
To cover a wide $T$ and $B$ range and avoid an experimental artifact in the measurement system, we used several different cryostats: a homemade $^3$He refrigerator at $T$ down to 0.5~K with magnetic fields up to 7~T, a $^3$He system (Heliox, Oxford) at $T$ down to 0.23~K with magnetic fields up to 17~T and a Physical Property Measurement System (PPMS, Quantum Design) with fields up to 9~T.
All leads were equipped with $RC$ filters ($R=1$~k$\Omega$ and $C=22$~nF).
In the DC measurements, a bias current was supplied by a DC current source (6220, Keithley) and the voltage was  measured with a nanovoltmeter (2182, Keithley). We measured the temperature dependence of resistivity for bias current $I$ by the current reversal method.
We confirmed that the contact resistance was $2-10$~k$\Omega$.

We performed magnetization measurements as a function of magnetic field and temperature using a Magnetic Property Measurement System (MPMS, Quantum Design) with applied magnetic fields up to 7~T and a temperature range of $2-300$~K.
The Ca$_2$RuO$_4$ nanofilm crystal powders was encapsulated.

\subsection*{Acknowledgements}

We thank Y. Tabata, K. Nakatsugawa, T. Kurosawa, K. Yamaya, S. Takayanagi, T. Matsuura, K. Inagaki, T. Honma, K. Ichimura, N. Matsunaga, N. Sakaguchi, and Y. Asano for experimental help and useful discussions.
This work was supported by JSPS KAKENHI (No.17K14326), Inamori Foundation, Iketani Science and Technology Foundation and Samco Science and Technology Foundation.

\subsection*{Author contributions}
H.N. and S.T. designed the experiments.
H.N., K.Y., Y.O., K.I., K.T. and T.N. synthesized the high-quality samples and analysed the crystal structure.
H.N., K.Y., Y.O. and K.I. performed the transport and magnetic measurements and analysed the data.
Y.K. performed the first-principle calculation.
K.N., T.A. and S.T. helped interpret the results.
H.N. drafted the manuscript.
All the authors read and approved the final manuscript.

\subsection*{Competing Interests}
The authors declare that they have no competing interests.

\newpage

\begin{figure}[t]
\begin{center}
\includegraphics[width=0.9\linewidth]{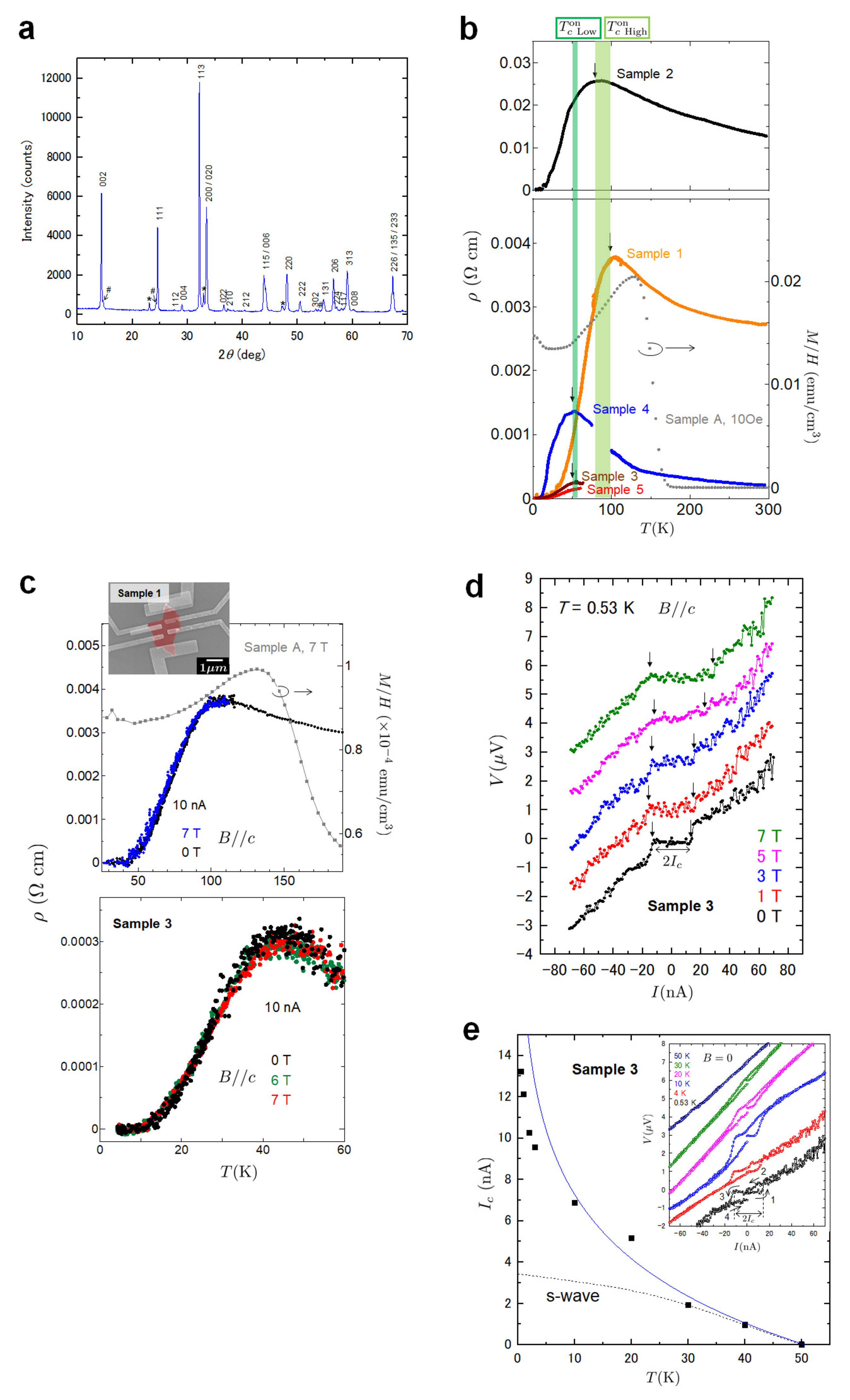}
\newpage
\end{center}
\end{figure}
\begin{figure}[t]
\begin{center}
\caption{
\textbf{Electric transport properties in Ca$_2$RuO$_4$ nanofilm crystals.}
(\textbf{a})Powder X-ray diffraction patterns at room temperature for Ca$_2$RuO$_4$.
(\textbf{b})Temperature dependence of the longitudinal resistivity $\rho$ for Ca$_2$RuO$_4$ nanofilm crystals.
Dependence of the magnetic susceptibility $M/H$ on temperature in an FC process of 10~Oe for powders weighing 5.2~mg (sample A).
(\textbf{c})Temperature dependence of $\rho$ at 10~nA for samples 1 and 3 when the magnetic field is applied parallel to the $c$ axis.
Dependence of $M/H$ on temperature in 7~T for sample A.
(\textbf{d})$I-V$ characteristics in various magnetic fields for sample 3.
Supercurrent was observed.
(\textbf{e})Dependence of the critical current $I_c$ on temperature in $B=0$.
The blue solid and dotted lines represent fitting results for the critical current in chiral $p$-wave and $s$-wave states, respectively.
The inset shows $I-V$ characteristics for various temperatures in sample 3.
Hysteresis behaviour was observed in the low bias current region.
Numbered arrows represent the bias current direction.
}
\label{figure1}
\end{center}
\end{figure}

\newpage

\begin{figure}[t]
\begin{center}
\includegraphics[width=0.85\linewidth]{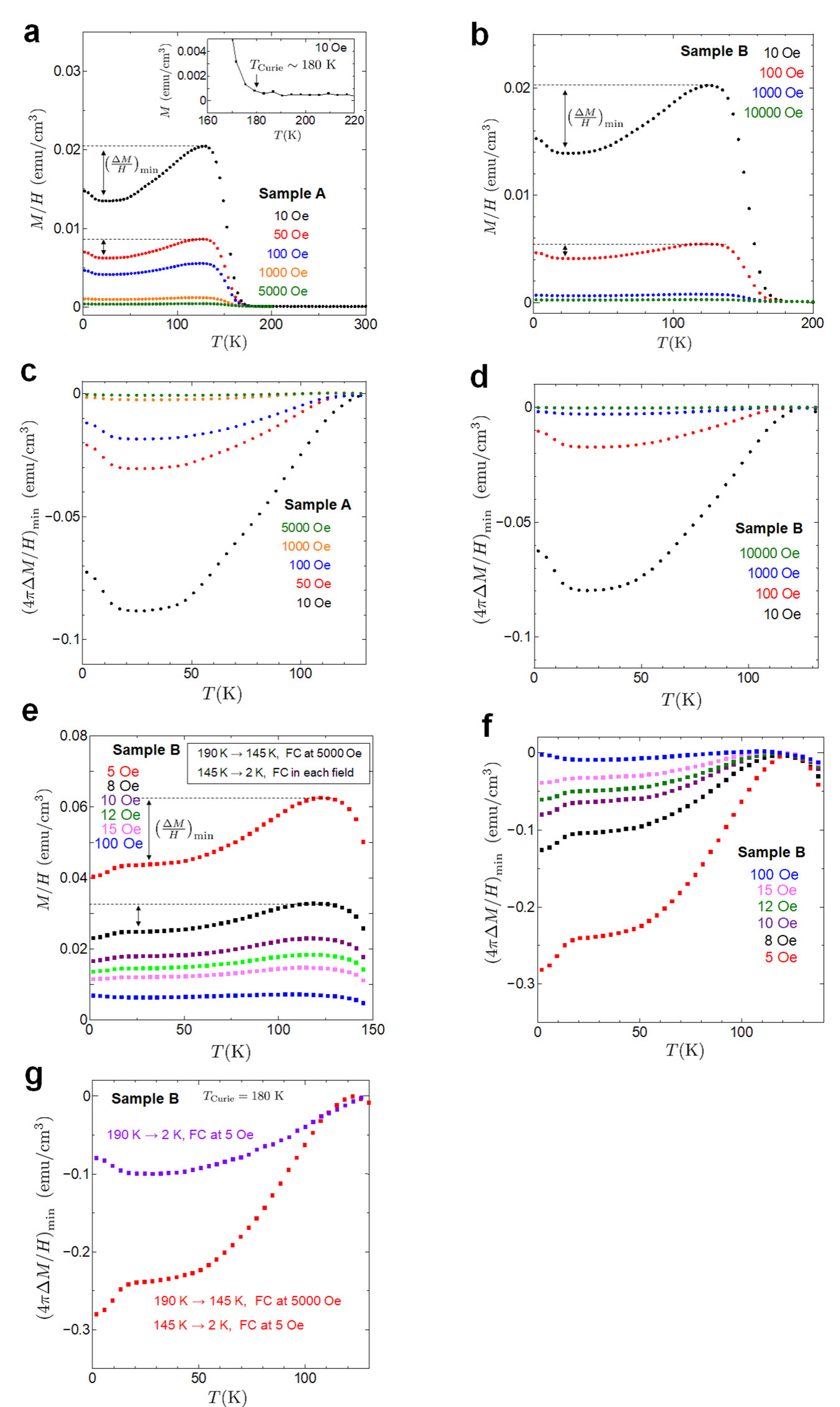}
\end{center}
\end{figure}
\begin{figure}[t]
\begin{center}

\caption{
\textbf{Magnetic measurements.}
(\textbf{a}) and (\textbf{b})Temperature dependence of magnetic susceptibility measured in the FC process for samples A and B. The inset shows an enlargement of the dependence of the magnetization on temperature around $T_{\mathrm{Curie}}$ measured at 10~Oe.
(\textbf{c}) and (\textbf{d})Dependence of diamagnetic susceptibility $\left(4\pi \Delta M/H\right)_{\mathrm{min}}$ on temperature for various magnetic fields in samples A and B.
(\textbf{e})Temperature dependence of magnetic susceptibility for FC from 145 to 2~K in various magnetic fields after cooling from 190 to 145~K while applying a magnetic field of 5000~Oe.
(\textbf{f})Dependence of diamagnetic susceptibility $\left(4\pi \Delta M/H\right)_{\mathrm{min}}$ on temperature for magnetic fields in sample B.
(\textbf{g})Comparison of diamagnetic susceptibility at 5~Oe in two different FC measurements.
The purple plot was measured by FC at 5~Oe from 190 to 2~K.
The red plot was measured by FC at 5~Oe from 145 to 2~K after cooling from 195 to 145~K while maintaining a magnetic field of 5000~Oe.
Larger diamagnetism was observed by aligning the direction of the ferromagnetic magnetization.
}
\label{figure2}
\end{center}
\end{figure}

\newpage

\begin{figure}[t]
\begin{center}
\includegraphics[width=0.9\linewidth]{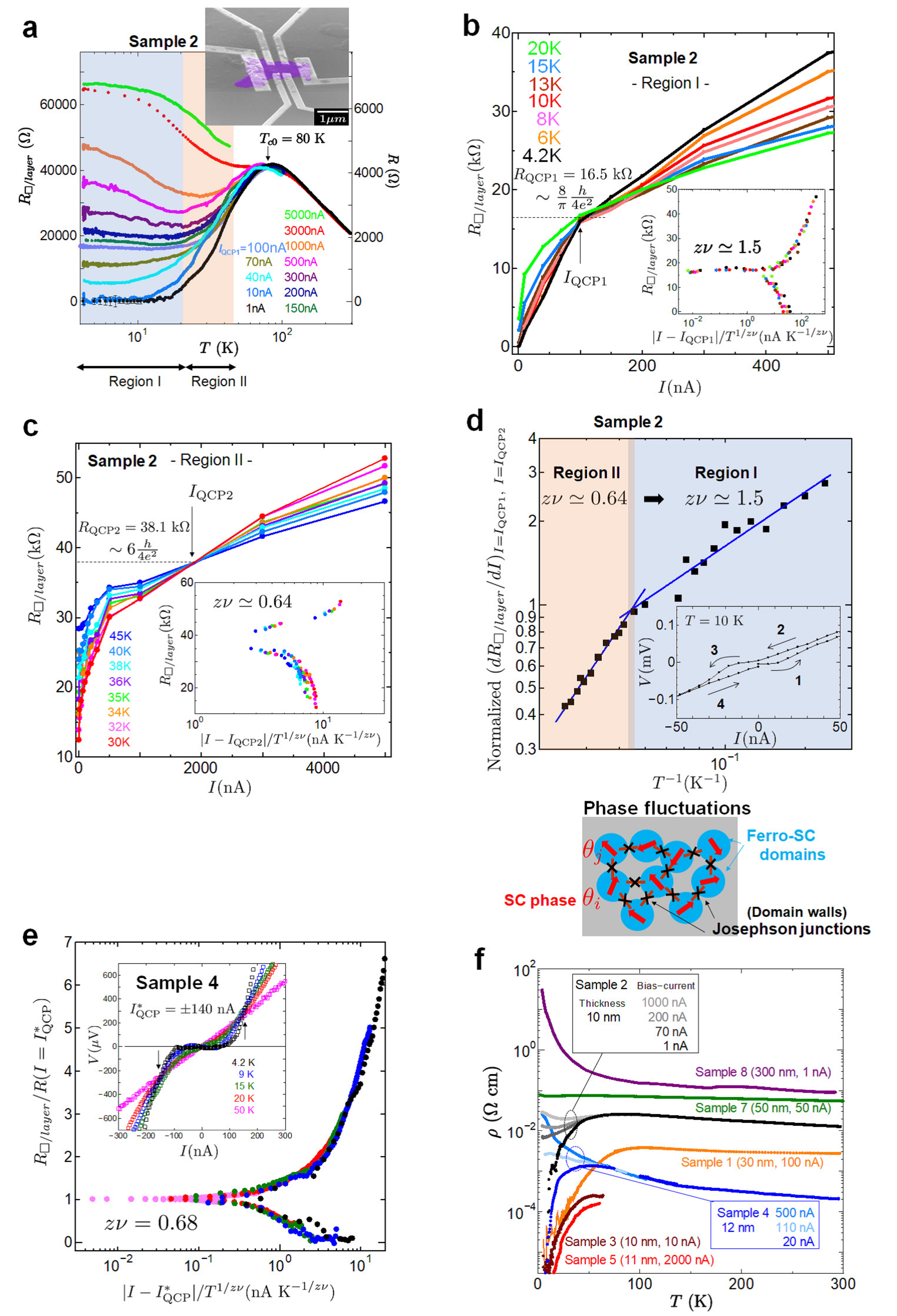}

\end{center}
\end{figure}
\begin{figure}[t]
\begin{center}

\caption{
\textbf{Superconductor-insulator transition.}
(\textbf{a})Temperature dependence of the sheet resistance $R_{\Box/layer}$ for sample 2 with a thickness of 10~nm.
(\textbf{b})Sheet resistance $R_{\Box/layer}$ as a function of bias current $I$ for temperatures ranging from 4.2 to 20~K.
The inset shows the scaling dependence of $R_{\Box/layer}$ as a function of the scaling variable $\left[c_{0}(I-I_{\mathrm{QCP1}})/T^{1/z\nu} \right]$ in the 4.2 to 20~K range where the values $I_{\mathrm{QCP1}}=100$~nA and $z\nu=1.5$ are used.
The critical sheet resistance $R_{\mathrm{QCP1}}$ was 16.5~k$\Omega \sim \frac{8}{\pi}\frac{h}{4e^2}$.
(\textbf{c})$R_{\Box/layer}$ as a function of $I$ for temperatures ranging from 30 to 45~K.
The inset shows the scaling plot of $R_{\Box/layer}$ as a function of $\left[c_{0}(I-I_{\mathrm{QCP2}})/T^{1/z\nu} \right]$ in the high temperature region, which corresponds to $z\nu \simeq 0.64$.
(\textbf{d})Normalized $(dR_{\Box/layer}/dI)$ as a function of $T^{-1}$.
Below $T_{c}^{\mathrm{onset}}$, two-stage current-induced critical points were observed.
The inset shows the hysteresis behaviour at 10~K in the low bias current region.
Schematic of superconducting domains coupled by tunneling junctions (lower inset).
(\textbf{e})Scaling dependence of $R_{\Box/layer}/R_{(I=I^*_{\mathrm{QCP}})}$ as a function of $|I-I^*_{\mathrm{QCP}}|/T^{1/z\nu}$ corresponding to $z\nu = 0.68$ from 4.2 to 50~K.
The scaling data were obtained from $I-V$ curves for sample 4 (inset).
(\textbf{f})Temperature dependence of the resistivity $\rho$ for Ca$_2$RuO$_4$ single crystals with different thicknesses.
Bias-current dependence is shown for samples 2 and 4.
}
\label{figure3}
\end{center}
\end{figure}

\begin{figure}[t]
\begin{center}
\includegraphics[width=0.5\linewidth]{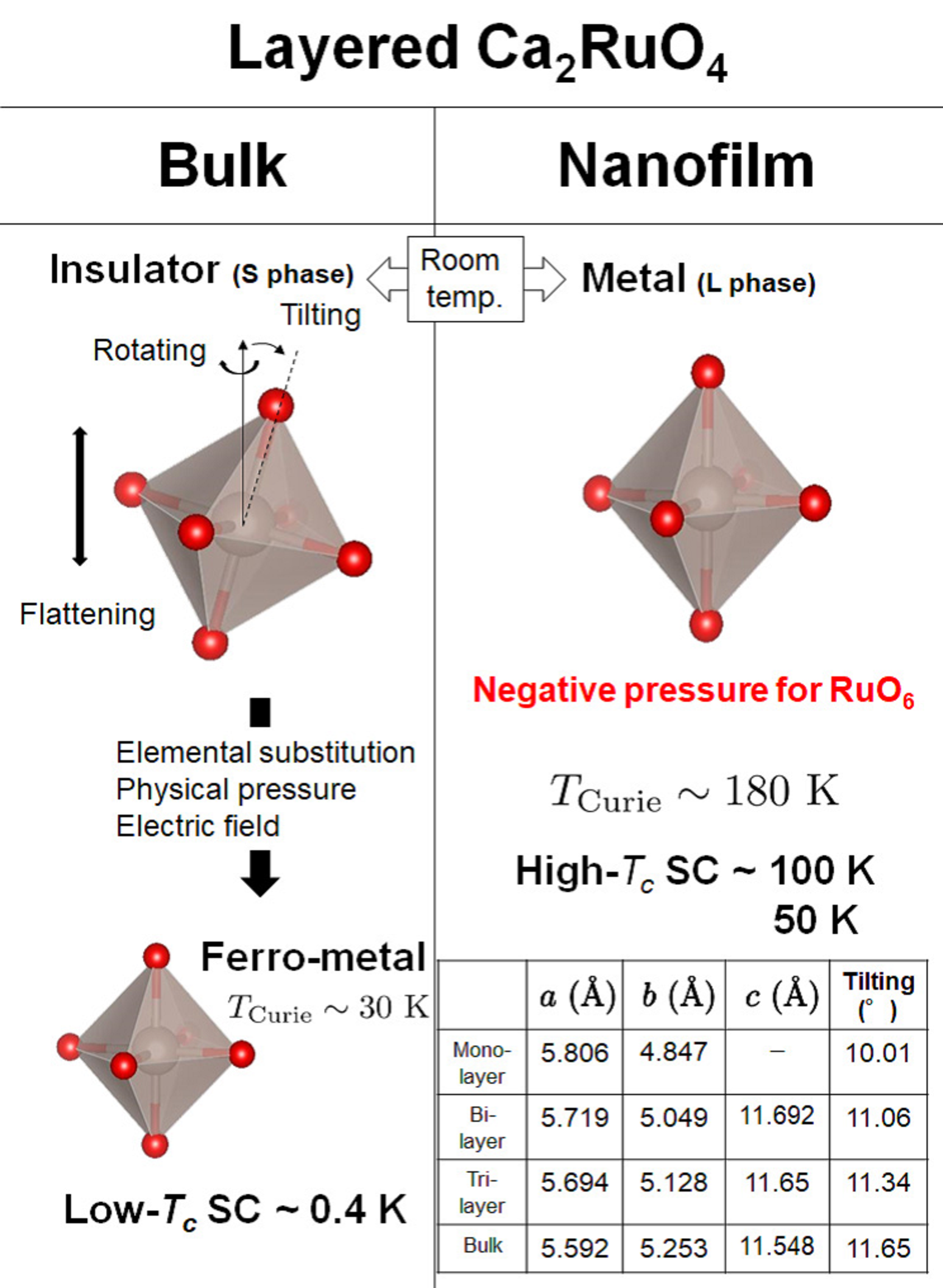}
\caption{
\textbf{Relationship between the ground state of Ca$_2$RuO$_4$ and the degree of
distortion of RuO$_6$ octahedra.}
Comparison of the ground states in bulk and Ca$_2$RuO$_4$ nanofilm crystals.
The results of our first-principle calculations for mono-, bi-, tri-layer and bulk Ca$_2$RuO$_4$ are shown in Table. The Ca$_2$RuO$_4$ thin film has weak distortions of RuO$_6$ compared with stacked bulk crystals.
}
\label{figure4}
\end{center}
\end{figure}

\end{document}